**Abstract**

Cold plasma technologies are an efficient approach to improve the germination properties of seeds, especially in a stacking configuration within a dielectric barrier device. In such dry atmospheric plasma priming process, we show that a helium-nitrogen plasma treatment of 20 min can reduce the median germination time of lentil seeds from 1420 min to 1145 minutes, i.e. a gain in vigor of 275 min (or +19.4 %). Considering that this result depends on the plasma-seed interaction and therefore on the contact surfaces between the seeds and the plasma, a topographic modeling of a 100 seeds-stack is performed in the dielectric barrier device. This model drives to the distinction between the seed-seed contact surfaces (276 contacts standing for a total area of 230.6 mm$^2$) and the seed-wall contact surfaces (134 contacts standing for a total area of 105.9 mm$^2$). It turns out that after a single plasma treatment, the outer envelope of each seed is 92% exposed to plasma: a value high enough to support the relevance of the plasma process but which also opens the way to process optimizations.

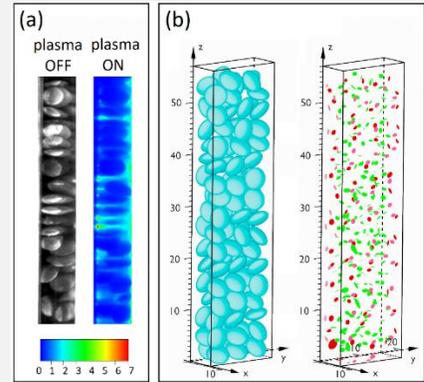

In this outlook, we propose to replace the single 20-minute plasma treatment by a "plasma sequence", i.e. a succession of shorter plasma treatments whose total duration remains 20 minutes. Between two successive plasma treatments, the seeds can follow either a trapping procedure (seeds in same positions and orientations) or a randomizing procedure (seeds in same positions but reoriented) or a stirring procedure (seeds vigorously shaken). As an example, a sequence of 10 plasma treatments (upon 2 minutes) separated by short stirring procedures leads to a gain in vigor as high as 405 min (+28.5 %) vs 275 min (+19.4 %) for a single plasma treatment of 20 min. We propose to understand these results by correlating the gain in vigor with the water uptake of the seeds (increase from 29% to 55%) and the wettability state of their coating (decrease of contact angle from 113.5° to about 38°).

# I. State of the art

## I.1. What is priming

In the life cycle of a seed plant, germination appears as the most critical developmental phase where the embryonic cells undergo a program3) to a highly active metabolic state (vigorous young seedling) [1]. As shown in Figure 1a, most mature seeds are found or stored in a dry state (S-phase) characterized by a water content typically lower than 10% and a resulting water potential as low as –100 to –350 MPa [2]. Sprinkling seeds with water drives to a water uptake mechanism sequenced into 3 canonical phases:

(i) The imbibition (I-phase) corresponds to a drastic increase in water content upon the first ten hours. The seeds exhibit limited biochemical activity, although light reactions and some oxidative processes are possible. The water impregnation of the seed is not homogeneous and depends on the nature of its constituents. Since protein is more hydrophilic than starch, the hydration of the protein-rich embryo is easier than the starchy endosperm [2]. The diffusion of water into the cells of dry seeds fragilizes their membranes, causing rapid leakage of solutes and low molecular weight metabolites (LMWM) into the surrounding imbibing solution. This phase is also characterized by DNA repair, mitochondria repair and synthesis of proteins using extant mRNA's [3].

(ii) The activation (A-phase) corresponds to a plateau, i.e. a period of limited or no water uptake although seeds undergo important physiological modifications, including mitochondria maturation, protein synthesis, storage reserve metabolism and production of specific enzymes [2]. As an example, the combined action of the proteasome and peptidases drive to the degradation of seed storage proteins.

(iii) The germination (G-phase) where a further increase in water uptake occurs and where the embryo axis elongates and breaks through the covering layers for the protrusion of the radicle [4].

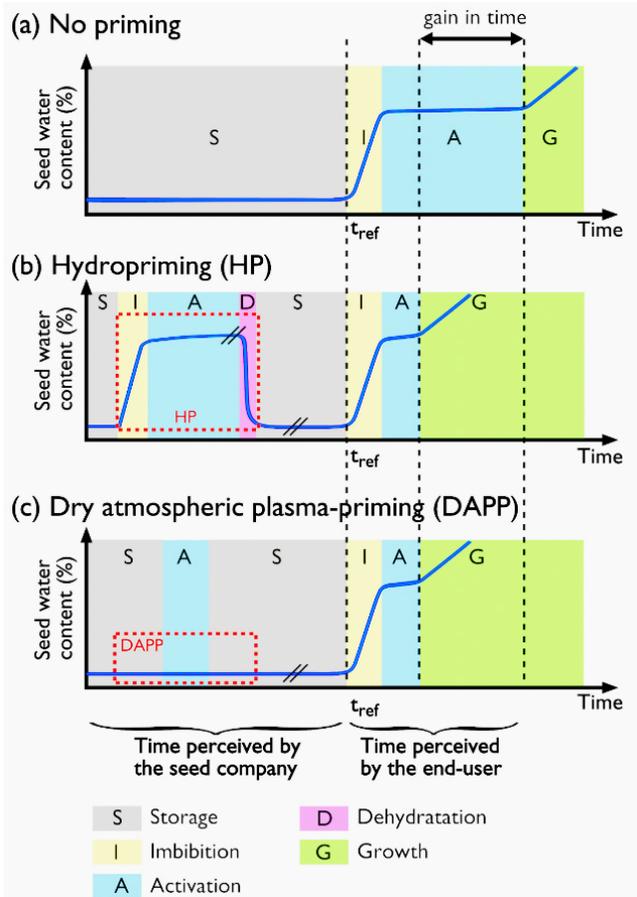

*Figure 1. Time variations of seed water content after contact with water (a) Timeline without priming technique, (b) Timeline including hydropriming process, (c) Timeline including dry atmospheric plasma-priming (DAPP). This figure is a summary diagram from [2], [8] and [10] adapted to plasma processes.*

Priming is a pre-sowing method commonly used by seed multiplier farmers, farmer seed enterprises and phytosanitary companies to improve the germinative properties of seeds, e.g. increasing the germination rate or improving the median germination time. This technique consists into partially hydrating seeds to initiate the metabolic activities of germination without bringing them to term, i.e. before reaching the radicle protrusion [5].

Among the various priming treatments carried out in routine, hydropriming appears as one of the simplest to implement. As sketched in Figure 1.b., seeds are soaked into water to achieve the imbibition (I-phase) and initiate the activation (A-phase) without fully completing it. Then, the seeds are re-dried (D-phase) to original moisture content to prevent the emergence of the radicle [6]. Upon this stage, orthodox seeds can be dried to water contents as low as 2-5% without risking any damage. Conversely, recalcitrant seeds cannot survive desiccation if water content is lower than 12-31% [7]. The main strength of this technique lies in the non-use of chemicals, making it both cheap and environment friendly. However, it should be pointed out that hydropriming suffers from several weaknesses: (i) its duration is relatively long, typically from 1 day to 1 week, while its multi-stage organization requires skilled workers, (ii) the water diffusion and absorption phenomena are uncontrolled owing to free water availability and to the affinity of seed tissues to water (iii) the non-uniform hydration of the seeds upon their imbibition phase result into an unsynchronized metabolic activation and therefore inhomogeneity in the appearance of radicles [8].

Alternatives to hydropriming include halopriming (seeds immersed in inorganic salts solutions like NaCl, $NaNO_3$, $MnSO_4$), thermopriming (seeds exposed to thermo-controlled environments), hormopriming (seeds imbibed in presence of growth regulators such as abscisic acid, auxins, gibberellins, …) as well as osmopriming [9]. That latter technique consists into soaking the seeds in a solution with a fixed osmotic potential to ensure a slower diffusion of water in the seed and therefore a better control of the first phases activating the germination process [10]. Thus, in the case of *R. pseudoacacia L.*, the effect of osmotic stress induced by the addition of increasing concentrations of $PEG_{6000}$ results in decreasing germination rates. As an example, soaking seeds at –0.2 MPa drives to germination rates twice lower than with pure water while soaking seeds at –1 MPa totally prevents the germination phenomenon [11].

## I.2. Dry atmospheric plasma priming (DAPP)

A different and innovative way to prime seeds relies on the utilization of cold atmospheric plasmas following a dry route. The so-called dry atmospheric plasma-priming (DAPP) could lead to similar performances as those obtained with conventional priming techniques with – as a major asset – a single-step treatment achieved on a much shorter processing time (< 1 hour). As already reported in the literature and as synthesized hereafter, DAPP can be achieved using various plasma reactors, some of them operating at low pressure (typically < 1 mbar) and others at atmospheric pressure.

Low-pressure RF plasma processes are successfully utilized to improve the germination of various seeds like *Moringa oleifera* seeds. Their exposure to an RF plasma of argon during 1 minute at 100 W and 2 Torr allows the increase in their germination by 20% and the subsequent shoot growth by more than 30 % [12]. Complementarily with these results, it has been shown that plasma can favor the diffusion of mineral elements in the coat of Quinoa seeds exposed to a capacitive RF reactor at 0.1 mbar, resulting in higher germination rate [13]. Low-pressure RF plasmas can also inactivate seed-borne bacteria showing a high pathogenicity. Hence, in the case of cabbage seeds artificially contaminated by *Xanthomonas campestris pv. campestris* (Xcc), Ono et al. demonstrates how plasma (13.56 MHz, 60 Pa, 100 W) can inactivate them and how such effect can be amplified by coupling oxygen plasma with a seeds' stirring process [14]. At higher frequencies, micro-wave plasmas are investigated to improve germination parameters. As an example, plasmas of helium operating at 3 GHz, 150 Pa and 80W can promote the germination rate of wheat (6.7% higher compared with control) as well as seedlings' growth parameters such as plant height and root length (+21.8 % and +11.0 % compared with control respectively) [15]. Interestingly and as detailed in the works of M. Soulier, hybrid plasma processes can be developed, for example to cumulate the advantages of high electronic densities (from RF plasma excitation) and uniform distribution of reactive species (from micro-wave plasma excitation) [16]. In the case of pepper seeds exposed to such RF-MW process, a significant increase in the germination rate is observed between control group (75 %) and plasma group (90 %). Another hybrid plasma process which deserves special attention is proposed by Ono et al.: seeds are exposed to a low-pressure RF plasma of air while being simultaneously placed under vibrating stirring. This mechanical effect drives to a deeper penetration of the active species from plasma and to higher pathogen inactivation rates, e.g. 10 times higher for *Xanthomonas campestris pv. campestris* on Cabbage seeds [17].

Although the biological effects resulting from these low-pressure plasma processes are important and reproducible, one major limitation remains their low processing capacity as well as the expensive cost of the low-pressure components (gauges, pumps, enclosures, consumables, …).

DAPP are also under study, in particular those operating in ambient air like corona and dielectric barrier devices (DBD). Several works show to which extent corona discharges can be utilized to decontaminate seeds from various pathogenic agents. In the case of broccoli seeds, microbial populations like aerobic bacteria, moulds, *B. cereus*, *E. coli* and *Salmonella spp.* are decreased following a 2-log factor only after 3 minutes of plasma exposure [18]. If simultaneous adverse effects like a decrease in germination rate is observed (from 78 % to 46 % for control and plasma groups respectively), other experiments show that, on the contrary, plasma corona processes can be engineered to purposely increase seeds' germination rate. Hence, the treatment of black soybean seeds to an air corona device for 2 minutes results into germination rates 3 times higher than those from control group [19]. In parallel to corona processes, dielectric barrier devices figure among the most promising reactors especially because seeds can be stacked in the interelectrode gap, thus offering the ability of treating large amounts of seeds. So far, most of these so-called seed-stacked dielectric barrier devices are supplied with a carrier gas (argon, helium) eventually mixed with a reactive gas. R. Molina et al. supply their DBD with helium to improve the germination of wheat seeds, while Ji et al. prefer molecular nitrogen to boost the slow germination rate of *Coriandum sativum* [20] [21]. Interestingly, if the interelectrode gap and the seeds characteristic diameters are small enough (typically < 4 mm), plasma can be directly ignited in ambient air, as shown in the works of Yi et al., making the process particularly attractive for industrials [22].

# II. Experimental setup, materials & methods

## II.1. Seeds

In this research work, lentil (*Lens culinaris*, from Radis & Capucine company) is used as an agronomic model for various reasons. First, the global annual production rate of lentils reaches 3.6 million tons, which makes this edible legume one of the most consumed in the world [23]. Second and despite this success, several (a)biotic stresses limit the growth and cultivation of lentils, hence the need to adopt new treatments instead of (or in addition to) the selection strategies which are currently being studied to improve their genotypes and adapt them to diverse agro-climatic conditions [24]. Third, lentil figures among the legumes requiring the shortest cooking time, typically 15 min, resulting in reduced nutrient losses (e.g. only 15% and 25% loss of thiamine and niacin respectively) [25]. Because they are rich in vegetable proteins, antioxidants, vitamins and minerals, lentils have many health benefits.

## II.2. Seed-stacked dielectric barrier device

The dielectric barrier device – whether seed-stacked or not – is constituted by a rectangular quartz tube (30 cm length, 6 mm × 15 mm inner section) platted by two aluminum electrodes. As sketched in Figure 2a, the resulting interelectrode gap (6 × 15 × 65 mm$^3$) can contain approximately 100 seeds of lentils (the unitary volume of one seed being 23.84 mm$^3$). The DBD is supplied with 6 kV in amplitude at a frequency of 500 Hz (sine form) using a high voltage generator composed of a function generator (ELC Annecy France, GF467AF) and a power amplifier (Crest Audio, 5500W,CC5500). It is supplied with a gas mixture of helium (1 slm) and molecular nitrogen (150 sccm) at atmospheric pressure. Before switching on the plasma, a purge procedure is achieved by injecting a He-N$_2$ gas mixture in the DBD to drastically reduce residual air background. In the Figures 2b (plasma off) and 2c (plasma on), the dielectric barrier is in contact with the upper horizontal plane of the seed stack while the counter electrode is in contact with its lower horizontal plane, consistently with Figure 2a.

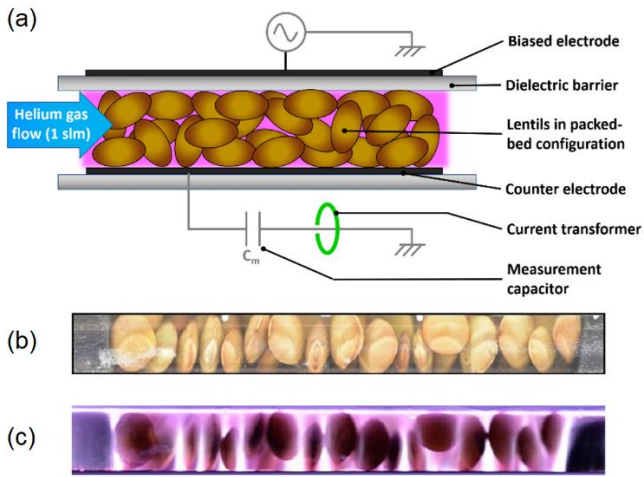

Figure 2. Sketch diagram of the seed-stacked dielectric barrier, (b) Picture of the seed-stacked DBD during $\tau_{off}$ and (c) during $\tau_{on}$.

## II.3. Sequences and procedures of dry atmospheric plasma priming

The seeds are exposed to the plasma following different sequences and procedures defined as follows. A sequence corresponds to a succession of N plasma treatments, keeping in mind that the duration of a sequence ($\tau_{sequence}$) corresponds to the duration of N single plasma treatments as written in equation {1}. Each plasma treatment can be broken down into an active phase where the seeds are effectively exposed to plasma during $\tau_{on}$ and a passive phase where seeds remain unexposed to plasma upon a timeout $\tau_{off}$ = 2 min (equation {2}). The Figures 2b and 2c illustrate the seed-stacked DBD during these two phases. For a given sequence, the plasma exposure times ($\tau_{on}$) have always the same value, hence explaining why equations {3} and {4} lead to equation {5}. Whatever the sequence, the total exposure time to plasma is $\tau_{pl}$ = 20 minutes as shown in equation {6}. The Figure 3a shows the 6 types of sequences studied in this work, considering N = 0, 1, 2, 4, 6 and 10 and the corresponding plasma exposure times $\tau_{on}$ = 0 min, 20 min, 10 min, 5 min, 3min20s and 2 min. For example, the sequence of N = 4 plasma treatments has a total duration of 28 minutes with 4 iterations of $\tau_{on}$ = 5 min and 4 iterations of $\tau_{off}$ = 2 min.

$$\tau_{sequence} = N . \tau_{treatment} \quad \{1\}$$
$$\tau_{treatment} = \tau_{on} + \tau_{off} \quad \{2\}$$
$$\tau_{sequence} = N . (\tau_{on} + \tau_{off}) \quad \{3\}$$
$$\tau_{sequence} = N . \tau_{on} + N \tau_{off} \quad \{4\}$$

$$\tau_{sequence} = \sum_{i=1}^{n} \tau_{on,i} + \sum_{i=1}^{n} \tau_{off,i} \quad \{5\}$$

$$\tau_{sequence} = \underbrace{\tau_{pl}}_{=20\,min} + \sum_{i=1}^{n} \tau_{off,i} \quad \{6\}$$

A procedure corresponds to the "mobility state" of the seeds upon the $\tau_{off}$ timeouts. Three 3 types of procedures are considered, as sketched in Figure 3b: (i) The "trapping procedure" where seeds' individual positions and orientations remain the same in the DBD upon each $\tau_{off}$ and therefore all along $\tau_{sequence}$, (ii) the "randomizing procedure" where a single reorientation of the seeds is performed in their stacked configuration so that their individual positions remain unchanged (iii) the "stirring procedure" where – during each $\tau_{off}$ – the seeds are extracted from the plasma-reactor, vigorously shaken during 1 minute in a sealed glass flask of 250 mL (mean stirring distance = 30 cm, mean frequency = 3 Hz) and reintroduced in the plasma reactor.

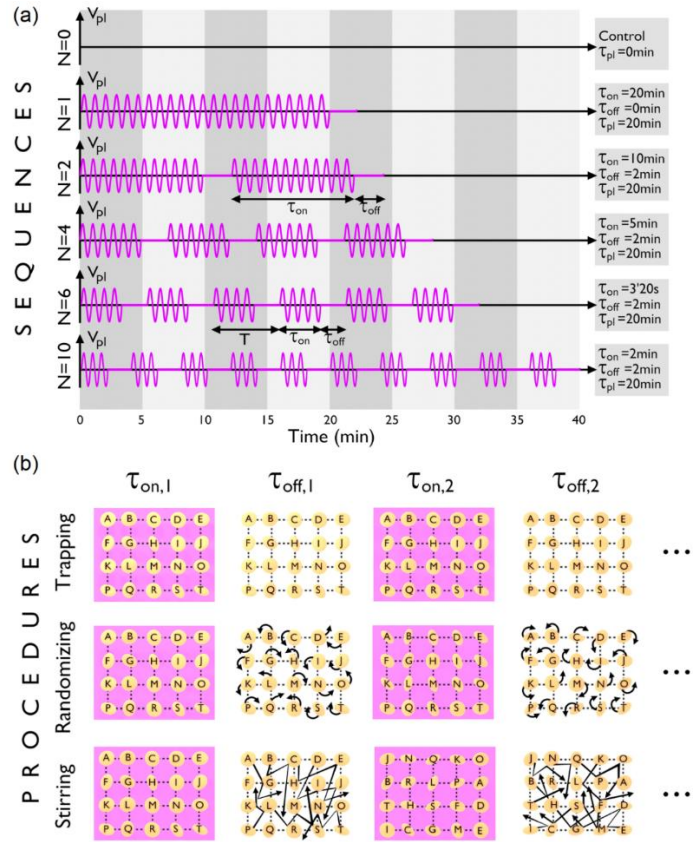

Figure 3. (a) Sequences of N treatments, each one including a plasma exposure phase ($\tau_{on}$) and a timeout phase ($\tau_{off}$=2 min. The total plasma exposure time $\tau_{pl}$ is always 20 min (b) Sketch diagrams illustrating the trapping, randomizing and stirring procedures carried out upon each $\tau_{off}$ phase. Seeds are referenced by letters to monitor their changes in locations.

## II.4. Diagnostics & Methods

### II.4.1. Electrical characterizations

The electrical characterization of plasma is achieved with an analog oscilloscope (Wavesurfer 3054, Teledyne LeCroy) and high voltage probes (Tektronix P6015A 1000:1, Teledyne LeCroy PPE 20 kV 1000:1, Teledyne LeCroy PP020 10:1). Current peaks are analyzed by placing a current transformer (Pearson company, model 2877) between the DBD counter-electrode and the ground, as sketched in Figure 2a. Then, a statistical analysis on peaks distributions is achieved using the Peak Analyzer toolbox of the Origin software. Baseline mode is set to 'Asymetric Least Squares Smoothing" while asymmetric factor, threshold, smoothing factor and number of iterations are set to 0.001, 0.02, 10 and 10 respectively. The electrical power deposited in the DBD is determined using the Lissajous method which relies on the addition of a capacitor ($C_m$ = 100 pF) between the counter-electrode and the ground.

Capacitance and resistance values of the seeds are measured using an LCRmeter (HM-8118 model from Rohde & Schwarz) so as to assess their apparent conductivity and permittivity.

### II.4.1. OES

Optical emission spectroscopy is utilized to identify the radiative species from plasma. The spectrometer (SR-750-B1-R model from Andor company) is equipped with an ICCD camera (Istar model) which operates in the Czerny Turner configuration. Its focal length is 750 mm while diffraction is achieved with a 1200 grooves.mm$^{-1}$ grating in the visible range. The following parameters are selected for all experiments: exposure time = 0.1 s, number of accumulations = 50, readout rate = 3 MHz, gate mode = 'fire only', intensification factor = 4000, insertion delay = 'ultra fast'.

### II.4.3. Temperature measurements

Temperature measurements of the seeds and of the walls' DBD reactor are performed using a thermal probe from TESTO company (model 826) that allows infrared temperature measurements (spectral range = 8-14 µm) as well as contact point measurements with an accuracy of 0.1 °C.

### II.4.4. Topographical modeling of the contact surfaces in the seed-stacked DBD

Solid modeling of the seed-stacked dielectric barrier device is achieved using Blender software (version 2.81). Blender allows to perform advanced real-time physics simulation using the Bullet Physics engine: a professional open-source library developed by E. Coumans et al. for collision detection, rigid body and soft body dynamics [26]. The library is free and the open-source software subject to the terms of the zlib License [27], [28]. Bullet Physics targets real-time and interactive use in games, visual effects, robotics and reinforcement learning. The physical properties of the Blender environment are utilized by default, in particular gravity with a value of -9.810 on the Z axis. The interelectrode gap of the DBD and a collection of 100 seeds are modelled by two types of mesh objects showing rigid body properties: the interelectrode gap is modeled by a parallelepiped mesh of 6 × 15 × 50 mm$^3$ while each seed is represented by an icosphere of level 6. This icosphere corresponds to a polyhedral ellipsoid characterized by its 20480 triangles and its three radii whose averages values ($\mu_1$ = 4.4, $\mu_2$ = 4.5 mm and $\mu_3$ = 2.1 mm) result from calculations performed on a sample of 30 real seeds of lentils. To make the model closest to the reality, each seed is made unique by slightly deforming it and by slightly changing its 3 radii. Their resizing is performed so that each of the 3 variables $\mu$ take 100 values according to a Gaussian distribution, centered around $\mu$ and included in the 2$\sigma$-range (95.4%). Among the 100-seeds collection, seeds present therefore the following typical dimensions: 4.4 ($\pm$1.0) × 4.5 ($\pm$1.0) × 2.1 ($\pm$1.0) mm$^3$. The collection and magnified views are shown in Figure 4.

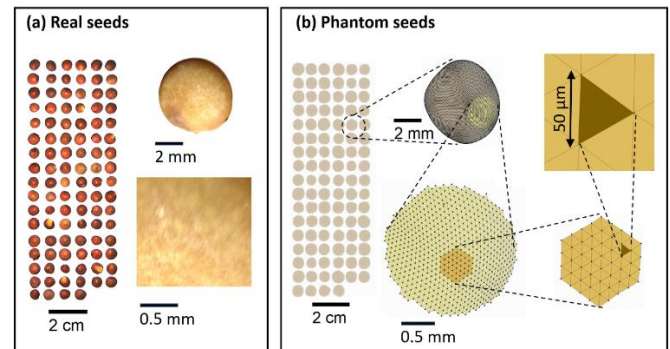

*Figure 4. (a) Collection of 100 real lentil seeds including magnification view on the outer surface of a seed, (b) Collection of 100 tridimensional phantom seeds modelled with Blender software including enlarged views on the tridimensional mesh. The elementary lengths and areas are 50µm and 1.25 × 10$^{-3}$ mm$^2$ respectively.*

### II.4.5. Drop shape analysis

Shape analysis of water drops deposited on seeds coating is achieved by measuring contact angles to evaluate the wettability state of seeds teguments. A drop shape analyzer engineered and assembled at laboratory is utilized to measure the water contact angles (WCA) following the Sessile drop method. All WCA measurements result from drops of milli-Q water, 10 µL in volume. Here, each WCA value is an average of 10 drops deposited on 10 different seeds.

### II.4.6. Water uptake

Water uptake is the parameter assessing the amount of water absorbed by a seed after its imbibition. It can be expressed in a dimensionless form as shown in equation {7}:

$$\xi(t) = 100 \times \frac{m(t) - m_0}{m_0} \quad \{7\}$$

where $m_0$ and $m(t)$ correspond to the mass of the seed sample in the native state and at a post-imbibition time t respectively. Here, one sample is composed of 50 seeds. The weighs are measured using an analytical balance (SARTORIUS company, model Entris 124i-1S) with readability of 0.1 mg and reproducibility ±0.1 mg.

### II.4.7. Germination assays

Germination assays are carried out in accordance with the recommendation of FAO [29]. The effects of plasma are systematically evaluated on 4 independent samples of 100 seeds, i.e. 400 seeds in total. The following procedure is followed:
(i) Utilization of organic cotton square pads (8cm × 8cm × 3mm) as test substrates placed in Petri dishes.
(ii) Soaking of each substrate with 10 mL of tap water whose chemical parameters are calculated from the sanitary control agency of the water distribution network in Paris and reported in Table 1 [30].
(iii) Distribution of each 100-seeds sample into 20-seeds batches per Petri dish. The seeds are evenly spaced from each other, at a distance of 2.5 times their diameter. Thus, the possible spread of fungal molds is avoided.
(iv) Storage of the seeds at 20 °C and RH = 40% in the dark throughout the germination follow-up.

| Parameter | Average |
|---|---|
| Turbidity | 0.4 NFU |
| Free chlorine | 0.1 mg($Cl_2$)/L |
| Conductivity | 590.8 µS/cm |
| pH | 7.6 |
| Iron | 9.4 µg/L |
| Nitrates | 47.8 mg/L |
| Ammonium | 0.0 mg/L |
| Escherichia Coli | 0 n/(100mL) |
| Hardness | 29.2 °f |
| Calcium | 108.1 mg/L |
| Bicarbonates | 295.0 mg/L |

*Table 1. Chemical parameters of the tap water used for seeds.*

# III. Results & discussion

## III.1. Characterization of the seed-stacked dielectric barrier device

Stacking seeds in a plasma reactor does not guarantee a systematic "plasma-seed" interaction, i.e. an action of the seeds on plasma properties correlated with an action of plasma on the seeds' biological properties. For example, a too low ionization degree of plasma will not drive to enhanced germination properties of the seeds. Likewise, placing only a dozen of seeds whose volume occupies 1% of the interelectrode volume will not significantly modify the plasma properties.

### III.1.1. Electrical characterizations

A simple way to verify the action of seeds on plasma is to evaluate the electrical properties of the DBD with/without seeds. On that purpose, the time profiles of the plasma voltage ($V_{pl}$) and plasma current ($I_{pl}$) are represented in Figures 5a and 5b respectively. If the absence/presence of seeds does not significantly change the profile of $V_{pl}$, a strong effect is observed on the peaks of current: their magnitude appears roughly twice lower if seeds are introduced in the plasma reactor. From an "equivalent electrical circuit" point of view, the stack of seeds corresponds to a resistor in parallel with a capacitor [34]. Its resistance value is so high that the electric charges passing from the electrode to the counter-electrode are never transferred through this stack; they are only transferred by the plasma, through the interstices. Therefore, the DBD voltage remains unchanged whether a stack of seeds is in the reactor (or not) and, where appropriate, regardless of the resistance value of this stack. Conversely, the presence of seeds in the reactor modifies the amplitude of the current peaks; these are more numerous and half as intense in amplitude. This is because when a streamer (plasma) propagates from the electrode to the counter-electrode, it encounters a certain number of dielectric obstacles (= seeds) in its path which locally modify the electric field. Depending on the topography of each seed, the electric field can then be more intense in some directions to the detriment of others, resulting in a more complex spatial distribution and to the generation of more current peaks of lower magnitude.

Owing to the measurement capacitor ($C_m$) placed between the counter-electrode and the ground (Figure 2), the Lissajous diagram of the DBD can be plotted, as shown in Figure 5c where $Q_{DBD}$ and $V_{DBD}$ are the electrical charge and the operating voltage applied between the two electrodes, respectively. The area comprised inside the closed curves are the same and correspond to a plasma power of 1.32 W (± 0.05 W) with/without seeds. However, depending on whether seeds are placed in the plasma reactor, Lissajous's figure shows two major discrepancies:

- The slopes of the AB edges are $\xi_{off}^{without} = 3.04 \, pF$ (without seeds) and $\xi_{off}^{with} = 4.95 \, pF$ (with seeds). Same values are obtained along the DC edges.
- The slopes of the DA edges are close so that $\xi_{on}^{without} \approx \xi_{on}^{with} \approx 37.5 \, pF$. Same values are obtained along the CD edges. Besides, these left and right edges exhibit discontinuities corresponding to the current peaks and therefore to the electrical charge of each plasma filament. These discontinuities are particularly visible when the DBD is not filled with seeds (black curve).

To clearly evidence the direct action of the seeds on the temporal distribution of $Q_{fil}$ (electrical charge of each individual plasma filament), a statistical analysis of the current peaks is achieved over 500 periods, hence covering an equivalent time of 1 s. The Figure 5d indicates that without seeds (black bars), the values of $Q_{fil}$ range from 1 nC to 50 nC with many filaments characterized by values in the 1-6 nC range. Stacking seeds in the DBD leads to a reorganization of the plasma filaments: the values of $Q_{fil}$ spread over a wider interval (from 1 nC to 87 nC) and are typically comprised between 8 nC and 20 nC. Also, the number of filaments is much without seeds: a maximum as high as $1.6 \times 10^5$ filaments is obtained at 4 nC versus only $1.0 \times 10^3$ filaments at 10 nC in the seed stack configuration. Finally, whatever the configuration (DBD with/without seeds), the total electrical charge remains the same over 1s and estimated to a value of 48.4 µC.

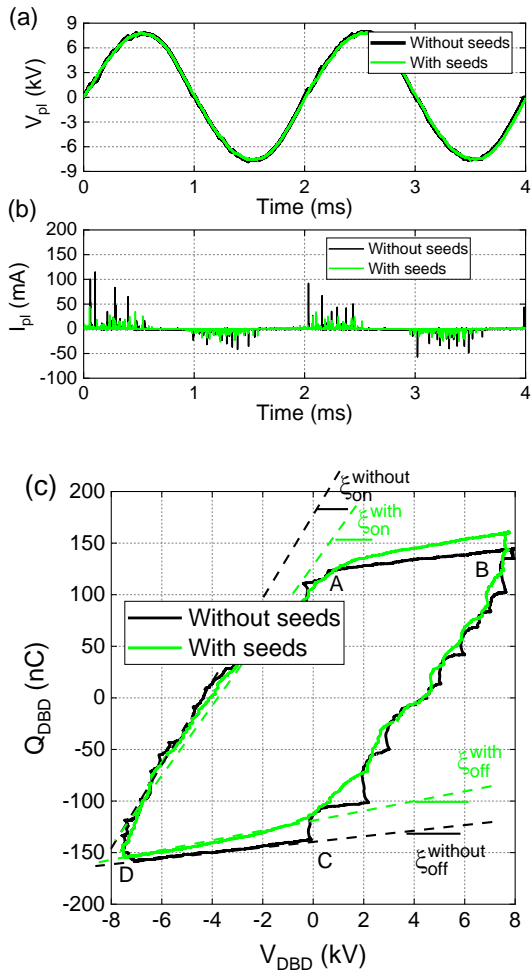

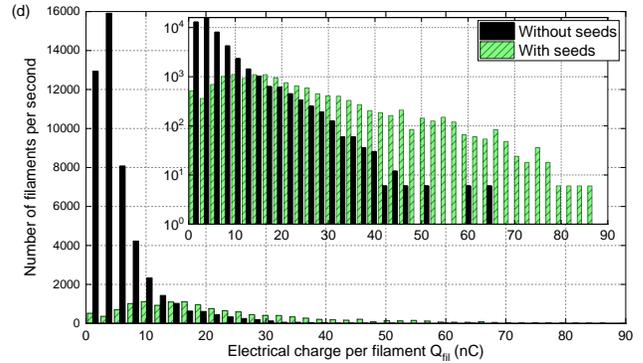

*Figure 5. (a) Time profile of the plasma voltage, (b) time profile of the plasma current, (c) Lissajous diagram of the DBD, (d) Distribution of the electrical charge per filament over 500 periods. In each plot, plasma is generated in helium gas (1 slm) mixed with $N_2$ (150 sccm), at 500 Hz and atmospheric pressure.*

### III.1.2. Identification of the active species in the plasma phase

If stacking seeds in the reactor modifies the electrical properties of plasma, one may expect similar modifications in its radiative properties. The Figure 6 (and its insets) shows the emission spectrum of the He-$N_2$ plasma for wavelengths comprised between 250 and 800 nm. Unsurprisingly, the helium peaks are detected (e.g. the one at 706.4 nm) as well as the bands of the second positive system of molecular nitrogen (e.g. the band headed at 337.1 nm). A low OH emissive band is also observed at 308.9 nm, probably due to traces of water vapor in the reactor and/or in the helium bottle. Overall, introducing seeds in the DBD drives to a significant decrease in all the emissive peaks and bands, with a rough factor of 10. This drop is to be correlated with the lower values of $Q_{fil}$ obtained in presence of seeds (Figure 5d).

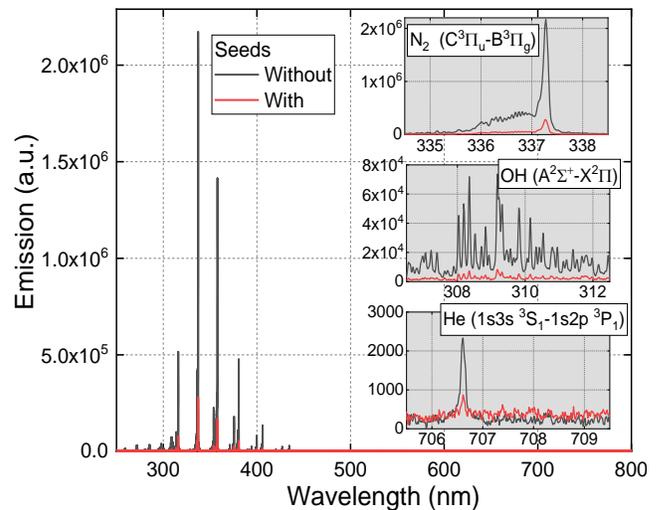

*Figure 6. Optical emissions spectrum of plasma generated in the DBD supplied in helium (1 slm) and molecular nitrogen (150 sccm), f=500Hz.*

More generally, the chemical composition of the plasma gas can play an important role on the germination properties of seeds, in particular on their vigor, as demonstrated in our previous work [31]. A plasma of helium with residual air background (Air/He <0.001) drives to the production of OH and O radicals that can contribute to improve the vigor parameter (gain in time close to +13%). The admixture of molecular oxygen to helium can however induce adverse biological effects owing to a too large production of ozone: at 150 sccm of $O_2$, the gain in vigor is only +9%. On the contrary, the addition of molecular nitrogen to helium helps greatly to reduce median germination times, through the production of reactive nitrogen species, with a gain in time close to +16%. For this reason, in this work, helium and molecular nitrogen flow rate were fixed at 1 slm and 150 sccm respectively (150 sccm corresponding to the maximum value of the flowmeter device).

### III.1.3. Characterization of the seeds stacked in the DBD reactor

The temperature of the walls' DBD reactor as well as of the seeds is measured during their exposure to plasma using the thermal probe from TESTO. Since a value of 33.4 °C (± 0.3 °C) is evidenced, one may wonder if the treatment does not risk slightly modifying the water content natively present in the seeds. The mass of a 100 seeds' stack is measured before and after its exposure to plasma during 20 min, resulting in the mass values of 3144.4 mg (± 0.8 mg) and 3127.5 mg (± 0.4 mg) respectively. As the relative variation of mass remains as low as $\frac{3144.4-3127.5}{3144.4} = 0.53\%$, one can reasonably consider that our treatment has negligible effects on seeds internal water organization.

The electrical capacitance and resistance of lentil seeds are also measured using an LCRmeter device to infer their apparent conductivity and relative permittivity. By assuming that the equivalent electrical model of a seed corresponds to a resistor in series with a capacitor, the LCRmeter provides the following resistance and capacitance values: $R_{1seed} = 12.564 \pm 0.704\ M\Omega$ and $C_{1seed} = 0.939 \pm 0.014\ pF$.

Electrical conductivity is estimated using equation {8}, considering L = 2.26 mm as the thickness of the seed (which is also the distance between the two LCR probes upon the measurement) while S = 1.589 $10^{-5}$ m² is the section of the seed. As a result, $\sigma_{1seed} = 1.131\ 10^{-5}\ S.m^{-1}$. Considering that the average mass of a single seed is 31.4 mg, the apparent conductivity is given by $\sigma_{app} = \frac{\sigma_{1seed}}{m_{1seed}} = 3.60\ \mu S.cm^{-1}.g^{-1}$. Complementarily, the relative permittivity of the seed can be estimated using equation {9} where $\varepsilon_0 = 8.854\ 10^{-12} F.m^{-1}$ is the vacuum permittivity. Even if this equation provides a rough approximation, we find $\varepsilon_r$ = 15.07: a value quite consistent with those in the literature [32] [33].

Although this article does not aim to understand the ins and outs of these two physical parameters and how they can influence the plasma properties, it appears that the values of $\sigma_{app}$ and $\varepsilon_r$ have not significantly changed before / after a 20 min plasma exposure. This is probably due to the native water content which is very low, and which cannot drive to a water loss higher than 1 %$_{wt}$.

$$\sigma_{1seed} = \frac{1}{R} \cdot \frac{L}{S} \quad \{8\}$$

$$\varepsilon_r = \frac{C_{1seed}}{\varepsilon_0} \cdot \frac{L}{S} \quad \{9\}$$

### III.1.4. Characterization of the contact surfaces within the seed-stacked DBD

If stacking seeds in the DBD modifies the electrical and optical properties of cold plasma, we will see that, in turn, the plasma can modify the seeds through the coating hydrophilization, their higher water uptake and their shorter median germination times (enhanced vigor). One can reasonably infer that the magnitude of these effects is directly correlated with the number and areas of the plasma-seed contact surfaces and therefore depends on several limiting factors:

(i) The existence of dead regions in the DBD where the gas is not ionized for various reasons (asymmetry of the electrodes, pressure gradients, electrodes slightly unparalleled, plasma generated by guided ionization waves, etc.)

(ii) As shown in the ICCD photograph in Figure 7a, the areas of the seed-seed contact surfaces ($a_{s/s}$) and seed-wall contact surfaces ($a_{s/w}$) are strongly related to the seeds' features, especially their geometry and stiffness:

   a. Concerning the geometry, it turns out that spherical seeds (e.g. cabbage, turnip) have very small $a_{s/s}$ contact areas unlike flattened seeds (e.g. pepper, carrot).
   b. Concerning the seed stiffness, two seeds with very hard seed coats or solid shells (e.g. sunflower seeds) cannot deform when placed in contact, resulting in a low contact surface area. Conversely, two soft seeds in contact can deform and share a larger contact surface area. This stiffness parameter depends also on exogeneous parameters like pressure, humidity and temperature. Thus, a plasma treatment whose gas temperature remains close to ambient temperature will not change the seeds thermoelastic properties while it will in the case of plasma exposures with gas temperature close to 60 °C.

Before assessing the total area of the seed-plasma contact surfaces ($A_{s/pl}$), one must evaluate the dimensional parameters of the seeds as unitary elements within the DBD. Considering that the three-dimensional representation of a seed is an ellipsoid, the calculation of its volume ($v_{seed}$) is given by equation {10} where $r_1$, $r_2$ and $r_3$ are the first, second and third radius of this ellipsoid.

Although the surface area of a seed ($a_{seed}$) can be accurately calculated using elliptic integrals (Jacobi integrals), the Thomsen formula seems more convenient as shown in equation {11}. The value of the p parameter is 1.6075, so that the maximum relative error in Thomsen's approximation is 1.061% at most [35]. Considering that the mean dimensions of a lentil seed are 4.43 × 4.55 × 2.11 mm³, the average volume and area of a single seed are $v_{seed} = 22.27 mm^3$ and $a_{seed} = 42.73\ mm^2$, respectively. Based on this information, $A_{s/pl}$ can be estimated using equation {11} where $a_{seed}$, $a_{s/s}$ and $a_{s/w}$ stand for the total area of a single seed, the area of a seed-seed contact surface and the area of a seed-wall contact surface, respectively. Similarly, in equation {12}, $A_{seeds}$, $A_{s/s}$ and $A_{s/w}$ represent the sums of these elementary areas inside the DBD. A solid modelling of the seed stack DBD is performed using Blender and the settings detailed in section II.3.3. so as to determine $A_{s/s}$ and $A_{s/w}$ (or $a_{s/s}$ and $a_{s/w}$) and then deduce $A_{s/pl}$.

The Figure 7b illustrates the resulting seed-seed contact surfaces (green color) and seed-wall contact surfaces (red color) modeled by Blender inside the 6 × 15 × 65 mm³ interelectrode gap considering a stack of 100 seeds. Unsurprisingly, the spatial distribution of these contact surfaces appears quite homogeneous as well as their orientations. For the sake of clarity, an enlarged view of the contact surfaces distributions is represented in Figure 7c including the values of their areas expressed in mm². The contour of each contact surface is perfectly smooth thanks to a large number of vertices and testifies a high mesh resolution. The Figure 7d reports the number of seed-seed and seed-wall contact surfaces as a function of their areas. These curves are fitted with Cauchy distributions (also known as Lorentzian functions) following equation {14}, where a* indicates the area for which the distribution peak is reached while γ stands for the half-width at half-maximum. The a* values are 0.61 mm² and 0.81 mm² for the seed-wall distribution and the seed-seed distribution respectively.

As reported in Table 2, global parameters can also be extracted from these distributions, especially their numbers: 134 seed-wall and 276 seed-seed contact surfaces are identified. The resulting number ratio of these contact surfaces is 0.48 and depends on seeds type a well as on the volume of the interelectrode gap. For a same type of seeds, a larger interelectrode gap will obviously drive to a smaller ratio while for a same interelectrode gap volume, smaller seeds will increase the value of this ratio. Considering that $a_{seed} = 42.73\ mm^2$ (equation {11}), the corresponding value of the seeds stack is $A_{seeds} = 4273\ mm^2$. Keeping in mind that the sum of all contact surface is $A_{S/S} + A_{S/W} = 336.5\ mm^2$, the proportion of $A_{seeds}$ which remains unexposed to plasma ($\eta_\%$) is deduced from equation {15} with a value of 7.9 %. In consequence, $A_{s/pl}$ is close to 92 %: a value large enough to guarantee the relevance of the plasma process for priming seeds but which, at the same time, opens the way to process optimizations aiming at improving the germinative parameters. This issue is addressed in the following sections (III.2 1 and III.3 2) through the implementation of sequences and procedures already introduced in section II.2 3.

$$v_{seed} = \frac{4}{3}\pi.[r_1 \times r_2 \times r_3]^3 \quad \{10\}$$

$$a_{seed} = 4\pi.\left[\frac{(r_1.r_2)^p + (r_1.r_3)^p + (r_2.r_3)^p}{3}\right]^{1/p} \quad \{11\}$$

$$A_{s/pl} = \sum a_{seed} - \sum a_{s/s} - \sum a_{s/w} \quad \{12\}$$

$$A_{s/pl} = A_{seeds} - A_{s/s} - A_{s/w} \quad \{13\}$$

$$f(x, a^*, \gamma) = \frac{1}{\pi\gamma.\left[1 + \left(\frac{x - a^*}{\gamma}\right)^2\right]} \quad \{14\}$$

$$\eta_\% = \frac{A_{s/s} + A_{s/w}}{A_{seeds}} \times 100 \quad \{15\}$$

$$\delta_\% = \frac{V_{seeds}}{V_{reactor}} \times 100 \quad \{16\}$$

|  | Types of contact surfaces | | |
| --- | --- | --- | --- |
|  | $A_{S/S}$ | $A_{S/W}$ | All |
| Number of contact surfaces | 276 | 134 | 409 |
| Sum of contact surfaces (mm²) | 230.6 | 105.9 | 336.5 |
| Average contact surface (mm²) | 0.84 (±0.02) | 0.79 (±0.07) | 0,82 (±0.03) |

*Table 2. Key numbers relating to the seed-to-seed ($a_{S/S}$) and seed-to-wall ($a_{S/W}$) contact surfaces.*

It should also be noted that in the DBD configuration, no dead volume (i.e. interstice without plasma) is observed, whatever the experimental conditions tested. On the contrary, all interstitial volumes contain emissive species, as shown in Figure 2c (photograph with exposure time = 500 ms) and Figure 7a (ICCD image with exposure time = 1 ms, 1000 accumulations). In this last image, the differences in emission intensity must be correlated with the size of the interstice volumes (and therefore with their respective optical depths) as well as with specific values of plasma density. This parameter results from the plasma micro-discharges whose spatial distribution is expected to be quasi-uniform. Since electric-field magnitude can reasonably be considered as reinforced in the immediate vicinity of the contact surfaces and then decreasing as one moves away from them, plasma interaction with the seeds is not uniform. Therefore, even if 92% of the seed surface is exposed to plasma, this exposure is probably not the same in intensity throughout the entire seed coating.

Finally, the volume occupancy rate ($\delta_\%$) of the seeds within the reaction volume (i.e., the volume placed in the interelectrode space) is a parameter which deserves our attention even if not directly linked to the contact surfaces. It is calculated according to equation {16} where $V_{seeds}$ and $V_{reactor}$ stand for the volume of the 100 seeds and for the reactor volume containing the seeds respectively. Considering that the average volume of a lentil seed is 23.84 mm$^3$ and that 100 of them are introduced in V$_{reactor}$ (6 × 15 × 65 mm$^3$), the occupancy rate is estimated at 40.7%. This value is significantly lower than the one calculated in our previous work where a value as high as 0.67 = 67 % is found because of the larger dimensions of the reactor (30 cm$^3$) [34].

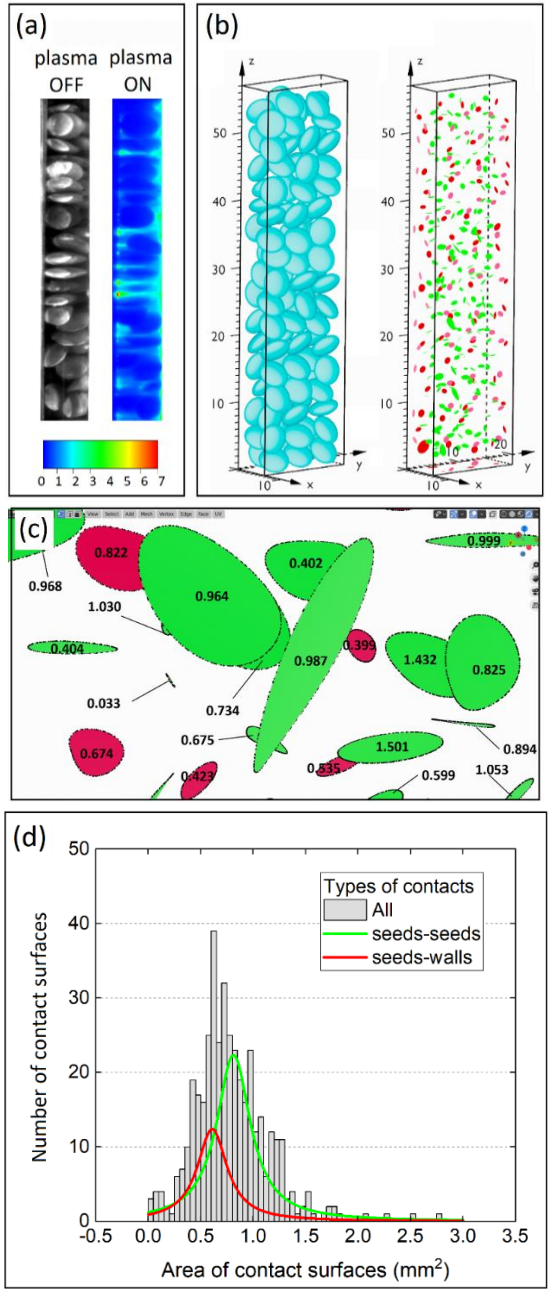

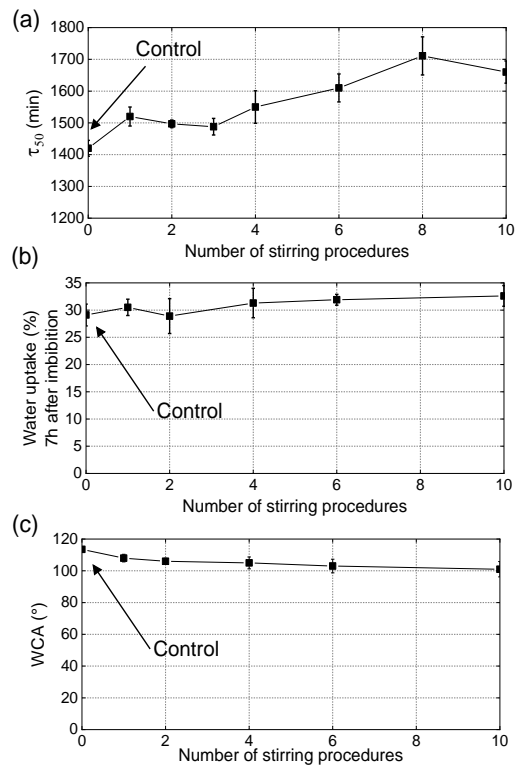

*Figure 7. (a) ICCD imaging pictures of seeds stacked in the dielectric barrier device during $\tau_{off}$ and $\tau_{on}$ phases, (b) Solid modeling of the seed-packing and its resulting contact surfaces, whether seed-seed (green, a$_{S/S}$) or seed-wall (red, a$_{S/W}$), (c) Snapshot of the seed-stacked DBD configuration simulated under Blender and showing the contact surface areas (mm$^2$) (d) Distribution of the contact surface areas in the seed-stacked DBD.*

## III.2. Seeds modifications resulting from plasma sequences and procedures

In the next experiments, seeds are always exposed to plasma during 20 minutes, consistently with Figure 3 and regardless the number of implemented sequences and procedures. As a reminder, we call sequence a succession of N plasma treatments and procedure a mobility state of the seeds upon the timeouts, i.e. trapping, randomizing or stirring procedure (section II.2.).

### III.2.1. Reference case: without plasma

Before estimating whether biological effects can be obtained by combining procedures and plasma sequences, it is first necessary to verify whether procedures alone are likely to induce such effects. In the case of trapping or randomizing procedures alone (i.e. no coupling with plasma), the biological effects would naturally be similar to those from the control group. However, the question arises for the stirring procedure where the shocks undergone by the seeds could modify the morphology of their tissues. Therefore, the lentil seeds are stirred according to the instructions detailed in section II.2. and N procedures of stirring are performed with N = 0, 1, 2, 3, 4, 6, 8 and 10 (2 minutes each, 100 seeds per 250 mL-sealed glass flask).

*Figure 8. Influence of the number of stirring procedures (a) on germination potential, (b) on water uptake of seeds 7h after their imbibition, (c) on contact angles of water drops deposited on seeds*

*coatings. Each stirring procedure lasts 2 minutes and is not combined with plasma.*

The influence of the stirring procedure – and more precisely the number of times it is repeated – is evaluated through 3 parameters described in Figure 8: the median germination time ($\tau_{50}$), the water uptake ($\xi$) and the water contact angles (WCA). In Figure 8a, a significant increase of $\tau_{50}$ is observed with a value increasing from 1420 min (control group) to more than 2000 min for N> 6. As a result, the stirring procedure slows down the germination vigor and seems inappropriate. In Figure 8b, the parameter $\xi$ measured 7 hours after imbibition shows values close to 30% regardless of N. Similarly, the WCA in Figure 7c are close to 110° for N comprised between 1 and 10. A very slight decrease seems visible but remains insufficient to conclude on a change in the wettability state. All these results clearly show that when the stirring procedure is carried out alone (i.e. not combined with plasm),, seeds' vigor is not improved.

### III.2.2. Effects on seeds vigor

The Figure 9 shows two germination curves of lentil seeds, one for the control group and the other for the plasma group (N = 1 treatment of 20 minutes). The median germination time ($\tau_{50}$) of a germination curve is deduced by the orthogonal projection on the time axis of 50% of the final germination rate. Here, $\tau_{50}$ reaches a value of 1400 min for the control group and only 1132 min for the plasma group. The gain in vigor ($G_{vigor}$) is estimated to 270 min and results from subtracting the two median germination times. The effect of plasma on the germination potential is not discussed here since a value as high as 100% is already reached for the control group.

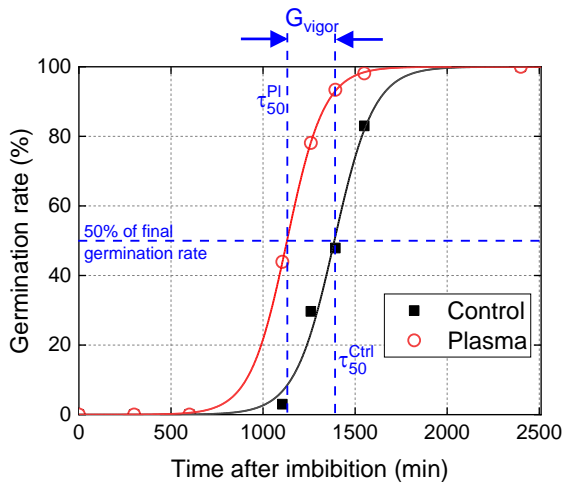

*Figure 9. Germination curves of lentil seeds either from control group or from plasma group (N=1, $\tau_{on} = 20\,min$)*

For a total plasma exposure $\tau_{pl}$ = 20 min, we propose now to verify if the fame value of Gvigor is obtained for different types of plasma sequences, i.e. for N varying between 1 and 10. The Figure 10 shows $G_{vigor}$ as a function of N considering the trapping (Figure 10a), randomizing (Figure 10b) and stirring (Figure 10c) procedures. Three important observations are noteworthy:

- A rise in $G_{vigor}$ is obtained whatever the procedure implemented within a sequence. This increase is especially important for N going from 1 to 2 and fades beyond, meaning that changing the position of the seeds only once guarantees a quasi-total exposure of the seeds to plasma. This observation can be supported on the point of a probabilistic view: while a single plasma treatment leaves $\eta_\%$ = 7.9 % of the outer seed coat unexposed to plasma, a sequence of 2 plasma treatments leaves $0.08^2$ = 6.4 × $10^{-3}$ = 0.64 % of surface unexposed to plasma. Then, a sequence of 3 plasma treatments would theoretically drive to $0.08^3$ = 5.12 × $10^{-5}$ = 0.05% of seeds surface unexposed to plasma. In consequence, implementing a higher number of plasma treatments (with shorter $\tau_{on}$ to maintain $\tau_{pl}$ = 20 min) would further randomize the seeds within the DBD but appears useless owing to the low value affected to the $\eta_\%$ parameter (here, only 7.9%). Conversely, other seed-stacked DBD characterized by higher $\eta_\%$ could require a higher number (N) of plasma treatments.
- The randomization procedure leads to values of $G_{vigor}$ that are higher than those of the trapping procedure where, as a reminder, the position and orientation of the seeds is unchanged during the timeouts of 2 min. While $G_{vigor}$ reaches a value of 275 min for a single treatment of 20 min, this same parameter peaks at around 340 min for N = 4, 6 and 10 (i.e. 5 min, 3min20s and 2 min respectively). Limiting randomization to N=2 or 3 seems appropriate in regard of the aforementioned probabilistic point of view.
- As seen in Figure 8a, the stirring procedure alone leads to an increase in the median germination time and has therefore no beneficial effect on germination. However, its sequencing with plasma drives exactly to the opposite phenomenon conclusion. Not only does the stirring procedure reduce median germination times (and therefore increase $G_{vigor}$ values), but it does also more effectively than the trapping and randomizing procedures. Furthermore, the $G_{vigor}$ values are even more important for sequences including high values of N: while $G_{vigor}$ is 275 min for a single treatment of 20 min, it goes to 345 min for 2 treatments ($\tau_{on}$ = 10 min) and to 405 min for 10 treatments ($\tau_{on}$ = 2 min). In all cases, the exposure time to the plasma remains the same but the cumulative duration of the stirrings increases, going from 1 min for N = 1 to 10 min for N = 10. It is considered that the stirring procedure enhances the effects of plasma through two phenomena. First, a stirring procedure performed after plasma could enhance the penetration of RONS from the outermost layers of the seeds to deeper layers as well as reveal areas of the seeds that were not exposed to plasma. Second, the stirring procedure generates shocks that could modify the seeds morphology, both on the outer layers (e.g. microcracks in the seed coat leading to faster imbibition of the seed) and in the bulk (e.g. reorganization of intercellular spaces).

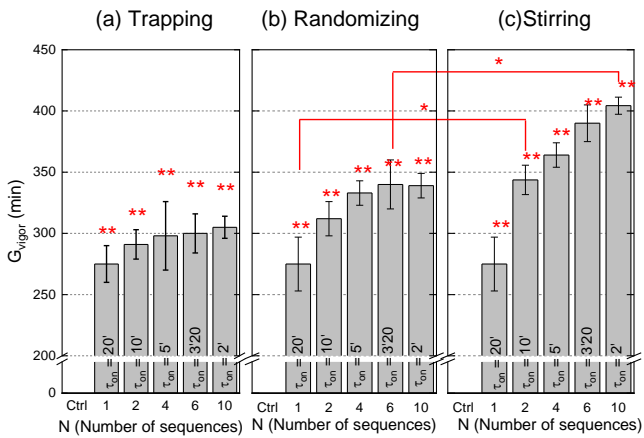

*Figure 10. Gain in vigor as a function of number of plasma sequences considering (a) trapping, (b) randomizing and (c) stirring procedures. For all the sequences, total plasma exposure time is $\tau_{pl}$ = 20 min and plasma timeout is $\tau_{off}$ = 2 min (* = 5% p-value, **=2.5% p-value)*

### III.2.3. Effects on seeds coating wettability

Although $\tau_{pl}$ = 20 minutes for all the experiments, great discrepancies are obtained on the resulting values of $G_{vigor}$. This biological parameter can be correlated with surface and core properties of the seed to better understand the underlying mechanisms. A simple way to proceed is to measure (i) the water contact angles (surface parameter) to deduce the integumentary wettability of the seeds and (ii) the imbibed masses (bulk parameter) to deduce seeds' water uptake.

The Figure 11a shows photographs of milli-Q water drops deposited on lentil seeds either unexposed or exposed to plasma. The effect of plasma is clearly evidenced through the spreading of the drop which attests a surface hydrophilization. This property can result from a chemical functionalization of the most superficial layers of the seed and/or from a modification of its physical properties, i.e. irreversible ablation or etching of these layers. The deformation of the drop is evaluated by measuring the WCA according to the Sessile drop method. The values of these WCAs are shown in Figure 11b for the N sequences and the three procedures. While the native WCA is estimated at 113.5°, a single 20 min plasma treatment lowers this value to 37.7°. It should be noted that an increase in N drives to a slight strengthening of this hydrophilic state only for the stirring or randomizing procedures but not for the trapping. Moreover, when N is fixed, the stirring always leads to WCA values lower than those of the randomizing, and the randomizing leads to values smaller than those of the trapping. We put forward that (i) the plasma allows the grafting of RONS in the form of chemical functions like O, OH, COOH and (ii) the stirring procedure favors the seeds' peeling, hence inducing an increase in surface roughness and therefore a larger apparent surface. The stirring procedure is also likely to favor the diffusion of reactive species towards deeper layers. As supported by optical emission spectroscopy, OH and $N_2^*$ species could be involved in the production and diffusion of RONS but other active species as well, especially the non-radiative ones.

Overall, the plasma sequences and procedures modify the seed wettability for values similar to those already obtained with the single treatment of 20 min. Since the discrepancies in the $G_{vigor}$ values greatly depend on the implementation of sequences and procedures, it seems unlikely that the modification of the surface properties explains the variations of $G_{vigor}$.

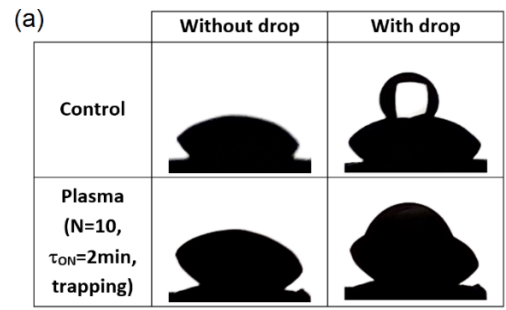

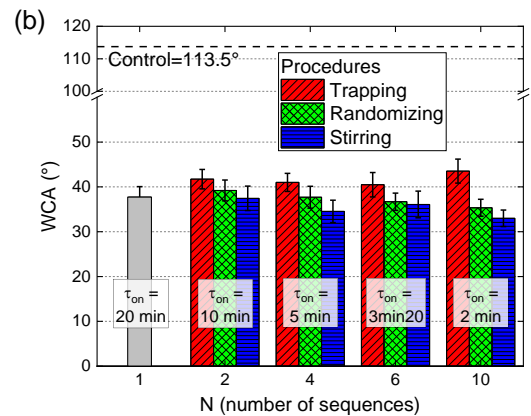

*Figure 11. (a) Photographic views of lentils seeds without/with drops (milli-Q water) deposited and analyzed following the Sessile drop method, (b) Contact angles of water drops deposited on seeds coating following several plasma sequences and procedures. For all cases, total plasma exposure time is $\tau_{pl}$ = 20 min and timeout is $\tau_{off}$ = 2 min.*

### III.2.4. Effects on seeds water uptake

The water uptake (ξ) defined by equation {7} is an indirect indicator of the seed's permeability as well as of the water retention capacity. Figure 12a represents ξ as a function of time for seeds originating either from the control group or from a plasma sequence including N=10 plasma treatments separated by stirring procedures. The discrepancy between the two curves is close to 20% from the first to the eleventh hour. The Figure 12b shows the value of ξ at t = 7h for the various plasma sequences and procedures described above. While ξ has a value of 29% for seeds in the control group, it turns around 40% for the seeds exposed to the single 20 min plasma treatment (N =1). Unsurprisingly, the coupling of plasma treatments with a trapping procedure does not change ξ whatever the values of N. Conversely, the randomization procedure slightly improves the water uptake and this effect is all the more reinforced as N is increased (e.g. ξ reaches a value of 44% for N = 10). Finally, the most remarkable effects are obtained with the stirring procedure since ξ = 48% for N = 2 and this value becomes as high as 55% for N = 10. All these results suggest that the stirring procedure substantially improves seeds water uptake but poorly affects the coating wettability.

It is interesting to note that the variations of ξ and WCA evolve in a coherent way: an increase in ξ (greater water absorption) being correlated with a decrease in WCA (hydrophilization). However, the variations of WCA are much smaller than those of ξ, probably because the analysis depth of the Drop Shape Analyser is typically limited to a few 10 nm (i.e. extreme surface). By their morphologies, seeds are rough and even porous structures. It is therefore entirely possible that reactive species can accumulate deep in the seed, where the Drop Shape Analyser can no longer detect them.

# IV. Conclusion

Stacking seeds in the DBD reactor drives to modifications of the plasma properties through (i) an increase in the number of filaments, with lower individual energies and a spreading of their distribution towards higher electrical charges, (ii) a decrease in the plasma emission through a decay in the He lines and molecular nitrogen bands. In turn, cold plasma in the DBD modifies the median germination times of seeds, driving to a gain in vigor of at least 4 hours in comparison with the control. The action of the seeds on the plasma properties and the simultaneous action of plasma on the seeds' germinative properties demonstrate the existence of a seed-plasma interaction. This interaction can be limited by dead regions in the DBD (resulting for example from non-paralleled electrodes) as well as by the number of seed-seed and seed-wall contact surfaces. In our configuration, the topographic modeling of the seed-stacked DBD predicts the existence of 409 contact surfaces, including 276 of seed-seed type and 134 of seed-wall type. The cumulative area of these contact surfaces is 336.5 mm$^2$ including 230.6 mm$^2$ for $A_{S/S}$ and 105.9 mm$^2$ for $A_{S/W}$. As a result, the proportion of seeds area exposed to plasma is estimated to 92% which supports the relevance of the process for technological transfer towards seed companies. The remaining 8 % can be exposed to plasma through a set of plasma sequences and procedures. In outlook, it could be interesting to verify this numerical estimate by performing non-intrusive experimental measurements based on microtomography analyzes on a synchrotron line.

For a same plasma exposure $\tau_{pl}$ = 20 min, a significant increase in $G_{vigor}$ is obtained through an increasing number of treatments performed over shorter $\tau_{on}$ and combined with procedures upon the timeouts ($\tau_{off}$ = 2 min). The three investigated procedures (trapping, randomizing and stirring) highlight the importance of the plasma-seed contact surface during the treatments. Indeed, $G_{vigor}$ can increase from 275 min (N = 1) to more than 405 min (N = 10, stirring procedure). The coating wettability and the water uptake have been measured before/after plasma treatment to correlate them with $G_{vigor}$. For the single-step plasma treatment of 20 min, WCA = 38°, against 113.5° for the control. Interestingly, hydrophilization may be slightly improved for specific sequences and procedures, especially those for a high number of randomizing or stirring procedures. With 10 stirring procedures, WCA can reach a value as low as 33°. We have also demonstrated a beneficial effect of plasma on seed water uptake: while a single plasma treatment increases ξ from 29% to 40%, the sequence-procedure combination improves this effect, especially for a high number of stirrings (ξ reaches a value as high as 56% for N=10). Performing plasma sequences and procedures seems an original and effective approach to optimize biological effects, as long as they have already been triggered by the plasma. In the case of lentil seeds stacked in the DBD, randomization can theoretically decrease $\eta_\%$ from 8% to 0.64 % (sequence of two treatments) and even down to 0.05 % (sequence of three treatments). The stirring procedure not only maximizes $A_{s/pl}$ (as in the randomization procedure) but also significantly improves vigor, probably due to better diffusion of RONS in the seed but also morphological modifications of the tissues whether outer (e.g. microcracks in the coating) or inner (e.g. intercellular spaces).

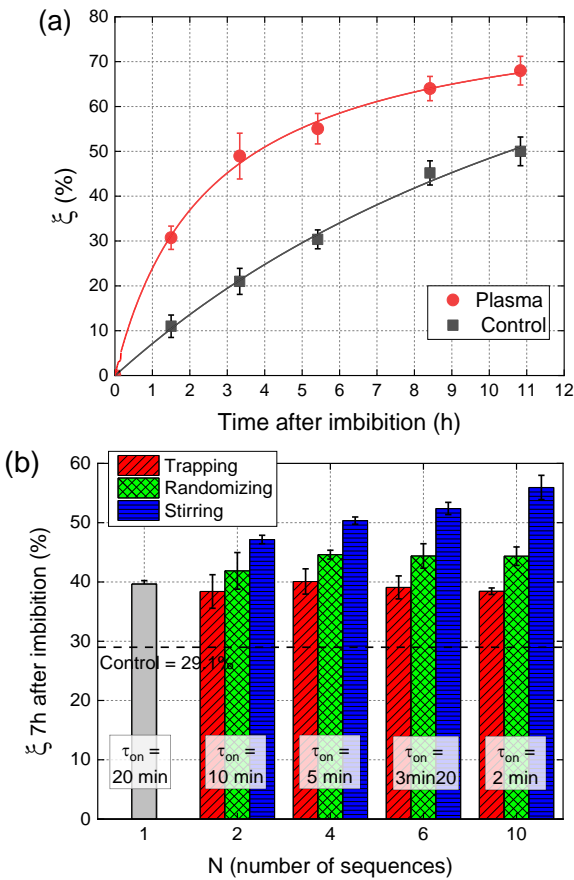

*Figure 12. (a) Water uptake of seeds as a function of tifsme after water imbibition comparing control to plasma (N=10, stirring), (b) Water uptake measured 7 hours after water imbibition for seeds exposed to plasma following different plasma sequences & procedures.*

Plasma sequence coupled to stirring procedure appears as a promising option to efficiently improve seeds vigor and potentially other germinative properties such as germination rate. Such coupling would deserve more experimental studies, especially to determine the optimal operation window of the shocks since a too low magnitude would have no effect while a too high magnitude could induce damages.

# V. Acknowledgements

This work is supported by grants from Région Ile-de-France (Sesame, Ref. No. 16016309) and ABIOMEDE platform through the Sorbonne Université Platform programme. This work is partly supported by French network No. GDR 2025 HAPPYBIO.# VI. References

[1] Marc Galland, Romain Huguet, Erwann Arc, Gwendal Cueff, Dominique Job, Loïc Rajjou, Dynamic proteomics emphasizes the importance of selective mRNA translation and protein turnover during Arabidopsis seed germination, Molecular & Cellular Proteomics, Vol. 13, No 1, 1-17 (2014)
http://dx.doi.org/10.1074/mcp.M113.032227

[2] Hudson T. Hartmann, Dale E. Kester, Fred T. Davies Jr., Robert L. Geneve, Hartmann and Kester's plant propagation principles and practices, Pearson Editor, 8th edition, ISBN-10 : 0135014492, ISBN-13 : 978-0135014493, 928 pages, Chapter 7 (2010)

[3] J. Derek Bewley, Seed germination and dormancy, The Plant Cell, Vol. 9, 1055-1 066 (1997)

[4] William E. Finch-Savage, Gerhard Leubner-Metzger, Seed dormancy and the control of germination, New Phytologist, Vol. 171, pp 501–523 (2006)
http://dx.doi.org/10.1111/j.1469-8137.2006.01787.x

[5] Pegah Moradi Dezfuli, Farzad Sharif-zadeh, Mohsen Janmohammadi, Influence of priming techniques on seed germination behavior of maize inbred lines (Zea mays L.), ARPN Journal of Agricultural and Biological Science, ISSN 1990-6145, Vol. 3, No. 3 (2008)

[6] Elien Lemmens, Lomme J. Deleu, Niels De Brier, Wannes L. De Man, Maurice De Proft, Els Prinsen, Jan A. Delcour, The Impact of Hydro-Priming and Osmo-Priming on Seedling Characteristics, Plant Hormone Concentrations, Activity of Selected Hydrolytic Enzymes, and Cell Wall and Phytate Hydrolysis in Sprouted Wheat (Triticum aestivum L.), ACS Omega, Vol. 4, 22089-22100 (2019)
http://dx.doi.org/10.1021/acsomega.9b03210

[7] E. H. Roberts, Predicting the storage life of seeds, Seed Sci. Technol., Vol. 1, 499-514 (1973)

[8] Stanley Lutts, Paolo Benincasa, Lukasz Wojtyla, Szymon Kubala S, Roberta Pace, Katzarina Lechowska, Muriel Quinet, Malgorzata Garnczarska, Seed Priming: New Comprehensive Approaches for an Old Empirical Technique, IntechOpen, Chapter 1, 46 pages (2016)
http://dx.doi.org/10.5772/64420

[9] Samira Hesabi, Saeed Vazan, Farid Golzardi, Investigation the effect of osmopriming and hydropriming on germination behaviour of alfalfa (Medicago sativa) and maize (Zea mays), International Journal of Biosciences, IJB, ISSN: 2220-6655 (Print) 2222-5234 (Online), Vol. 5, No. 6, 182-188 (2014)
http://dx.doi.org/10.12692/ijb/5.6.182-188

[10] Giuseppe Di Girolamo, Lorenzo Barbanti, Treatment conditions and biochemical processes influencing seed priming effectiveness, Italian Journal of Agronomy, Vol. 7, No 25, 178-188 (2012)
http://dx.doi.org/10.4081/ija.2012.25

[11] Mohamed Toumi, Selma Barris, Mohamed Seghiri, Houmam Cheriguene, Fatiha Aid, Effet de plusieurs méthodes de scarification et du stress osmotique sur la germination des graines de Robinia pseudoacacia L., Effect of several methods of scarification and osmotic stress on seed germination of Robinia pseudoacacia L., C. R. Biologies, Vol. 340, 264-270 (2017)
http://dx.doi.org/10.1016/j.crvi.2017.02.002

[12] Nagia Dawood, Effect of RF plasma on Moringa seeds germination and growth, Journal of Taibah University for Science, Vol. 14, No. 1, 279-284 (2020)
http://dx.doi.org/10.1080/16583655.2020.1713570

[13] A. Gómez-Ramírez, C. López-Santos, M. Cantos, J. L. García, R. Molina, J. Cotrino, J. P. Espinós, A. R. González-Elipe, Surface chemistry and germination improvement of Quinoa seeds subjected to plasma activation, Scientific Reports, Vol. 7, 5924 (2017)
http://dx.doi.org/10.1038/s41598-017-06164-514

*T. Dufour, Q. Guttierrez, J. Phys. D: Appl. Phys., 54, 505202 (16pp), 2021* https://doi.org/10.1088/1361-6463/ac25af